# Event-driven Implicit Authentication for Mobile Access Control

Feng Yao, Suleiman Y. Yerima, BooJoong Kang, Sakir Sezer
Centre for Secure Information Technologies,
Queen's University Belfast
Belfast, Northern Ireland, UK
fyao02@qub.ac.uk, s.yerima@qub.ac.uk, B.Kang@qub.ac.uk, s.sezer@ecit.qub.ac.uk

*Abstract*—**In order to protect user privacy on mobile devices, an event-driven implicit authentication scheme is proposed in this paper. Several methods of utilizing the scheme for recognizing legitimate user behavior are investigated. The investigated methods compute an aggregate score and a threshold in real-time to determine the trust level of the current user using real data derived from user interaction with the device. The proposed scheme is designed to: operate completely in the background, require minimal training period, enable high user recognition rate for implicit authentication, and prompt detection of abnormal activity that can be used to trigger explicitly authenticated access control. In this paper, we investigate threshold computation through standard deviation and EWMA (exponentially weighted moving average) based algorithms. The result of extensive experiments on user data collected over a period of several weeks from an Android phone indicates that our proposed approach is feasible and effective for lightweight real-time implicit authentication on mobile smartphones.**

## I. INTRODUCTION

Mobile devices have become indispensable to daily life, providing convenient access to online services many of which involve monetary transactions and storage or transfer of sensitive personal data. As a result, they are increasingly becoming attractive targets to various malicious threat actors that can compromise the security and privacy of data accessed from and/or stored on such devices. Since mobile devices are particularly vulnerable to loss or physical theft, it is important to have effective access control and strong protective measures that secure the devices at all times. Access control is usually enabled by authentication of the user (i.e. the system should verify that the user is legitimate).

Traditionally, user authentication has been based on three categories: 'what you know', 'what you have' and 'what you are'. Password based authentication is a simple example based on 'what you know'. It is renowned for its simplicity and ease of use. However, a recent survey [1] revealed that users choose convenience over password-based security. Preference over weak password and reuse of same password has become a common phenomenon and users prefer to keep themselves perpetually logged-in to sensitive online accounts and applications unless required by the application to log in every time, which makes user information vulnerable to theft (despite the use of passwords). Authentication based on 'what you have', such as adding a smart card reader on the phone or OTP (one time password) tokens is highly inconvenient and easily lost. Authentication based on 'what you are' is summarized in [2] which includes biometric features such as face and iris, fingerprint, voice/speech. However, biometric-based authentication schemes are computationally expensive for memory and power constrained devices.

Recently, *implicit authentication* has been investigated as an alternative to the traditional authentication methods. For example, [3-6] proposed schemes based on user behavior patterns or context. Although the results are promising, maintaining a balance between accuracy, adaptiveness, and practical feasibility is still an unresolved challenge. Hence, in this paper we propose and investigate an event-driven scheme that incorporates user-behavior awareness through data from everyday interaction with the phone. In order to provide real-time, lightweight and transparent operation in practical real-life scenarios, we consider easily derived user behavior features for user profiling and employ aggregate scoring and threshold computation for the implicit authentication scheme. This can then be used to activate or trigger implicit authentication (e.g. password, pin, voice, etc.) when there is a deviation from the usual or routine user behavior. Compared to previous works, important contributions in our proposed and approach include:

1) *Transparent operation*: No manual interaction or feedback from the user is required.
2) *Adaptivity*: Designed to adapt to shifts in user behavior, with the method employing EWMA-based threshold computation requiring no training phase.
3) *Event-driven*: No requirement for continuously running background services unlike many existing authentication schemes.

In the rest of the paper, Section II reviews related work in mobile user behavior-based authentication while Section III describes our proposed authentication scheme. Section IV presents the evaluation of the scheme using real captured data from an Android device. The paper is concluded in section V.

## II. RELATED WORK

Related work on protecting data on mobile devices based on behavior or context can be found in the current literature.

TABLE I. COMPARING DIFFERENT FEATURES OF OUR PROPOSED SCHEME WITH PREVIOUS WORK

| | Adaptivity | User Interaction | Threshold | On-device operation |
|---|---|---|---|---|
| **Proposed scheme in this paper** | High | None | Dynamic | Yes |
| **Implicit auth. [3]** | Low | None | Static | Yes |
| **Progressive [4]** | Medium | Yes | Static | No |
| **CASA [5]** | Medium | Yes | -- | Yes |
| **Intuitive [6]** | High | Yes | Static | Yes |
| **Data-driven [7]** | Medium | None | Static | Yes |

For example, a scheme called progressive authentication is proposed in [4] which utilizes face, voice recognition and other features to determine the level of confidence in a user's authenticity. Based on that level, access to public, private or confidential content is granted. Gait recognition is utilized in [8] to detect whether the device is being used by the rightful owner or not. However, the accuracy of result can be affected by footwear, ground surface, carrying load and injuries which can alter a person's usual walking style. [9] enhances password patterns robustness with an additional touch screen security layer which is called touch screen pattern. Authenticating mobile device users through keystroke analysis is investigated in [10,11]. In order to provide reliable scores, keystroke based authentication usually requires a rather long training phase. If the device uses a soft input method instead of hardware keyboard, it will further complicate the issue.

A type of implicit authentication scheme is proposed in [3] which creates a learning and recognition pattern by collecting and analyzing the daily behavior (phone calls, SMS history, browser history, network information etc.) of the user. The implicit authentication scheme in [3] scores recent user behavior based on the probability density function and requires a large set of user data as training dataset. In contrast, our proposed scheme employs a different approach that does not rely on a user data training set despite employing a set of similar features (utilized in a different way). Moreover, our scheme unlike [3], is inherently capable of adapting automatically to shifts in user behavior.

Another behavior based authentication scheme is proposed in [7] which utilizes probability density function to model user behavior temporally and spatially. This scheme is capable of automatically switching from training to deployment mode and determining a suitable detection threshold. It is also able to activate a retraining module in reaction to changes in user pattern. However, this scheme samples the sensor data 1440 times a day for both temporal model and spatial model, which means it frequently requires GPS or mobile network location (if GPS is not available) functions which could quickly drain the phone battery. As users frequently disable these services, it may not be a very practical solution.

A context-aware scalable authentication scheme is proposed in [5] which is designed to reduce active authentication on the mobile device with the inference of surrounding context. This scheme infers the legitimacy of the user by implementing a probabilistic framework for dynamically selecting an active authentication scheme that satisfies a specified security requirement given passive factors (location).

Another context profiling framework is proposed in [6] which utilizes context variables like GPS readings, WIFI access point and Bluetooth devices. This framework is able to estimate the familiarity and safety of a context based on the context variables and use it to dynamically configure security policies. This scheme also allows user to provide feedback to calibrate the perceived safety of context.

In Table I, a comparison of our proposed scheme with previous ones is summarized. As shown in the table, with the exception of the 'Progressive' scheme, all others are operated on-device. Thus, potential privacy leakage is prevented. For user interaction, fewer interactions means more convenience for users. Non-interaction also implies transparency and removes the need for complex user configuration, which can also introduce errors to the proper functioning of the authentication process.

As for threshold and adaptability, the authentication scheme proposed in this paper computes (decision) threshold(s) dynamically, unlike the other schemes. The 'intuitive authentication scheme' [6] uses a static threshold to determine the safety level, but it updates CoI (context of interest) with every new observation and uses WIFI access point and Bluetooth devices as context variables. Therefore these two authentication schemes are capable of effectively dealing with the change in user behavior patterns. Data-driven authentication [7] adopts a static threshold and conducts retraining process when the behavior drift has been detected, which means this authentication scheme could leave mobile devices unprotected during the retraining phase. Hence, data-driven authentication is deemed to be able to moderately deal with user behavior change. The 'implicit authentication' scheme in [3] builds a user model based on a training dataset and adopts a static threshold. A static threshold will limit its ability to detect drifts in user behavior patterns which could occur due to events like long distance travel or move to another city for example.

III. PROPOSED SCHEME

The user authentication scheme proposed in this paper models the user profile through the data collected from the user behavior. A scoring algorithm is applied to compute the level of trust of the current user. A decision threshold is adaptively computed below which an explicit authentication process can be triggered or activated to provide access control. Our scheme is aimed at operation without dependence on a user training set as much as possible. Thus, EWMA-based decision threshold computation is adopted as this enables zero (EWMA_Direct) or minimal (EWMA_SD_Block) training phase.

*A. Feature-based user profile modelling*

The proposed approach utilizes features derived from events depicting everyday user routines or patterns of behavior. It aims to use easily obtainable, readily available and commonly found data that can be extracted from most phones in order to construct a user's profile from routine phone usage. Hence, we consider the following as sources of data for our implicit authentication scheme:

1) Incoming/outgoing SMS
2) Incoming/outgoing phone call
3) Browser history
4) WIFI history
5) App usage patterns
6) Location
7) Accelerometer measurements

The experiments presented in this paper are based on the first four as sources of data. We plan to extend the scheme to include data from 5), 6), and 7) as well as other sources in the future to further enrich the user profiling capability.

The workflow of proposed authentication scheme is presented in Fig. 1. Every time a feature-related event occurs, the event is captured by the event monitoring module. After that, the scoring module computes a score for the associated feature (e.g. if there is an incoming/outgoing SMS, the score for the incoming/outgoing SMS is calculated), and then computes a new aggregate score. Then the threshold computation module computes the threshold from the score(s). The decision module compares the current aggregate score and threshold to determine whether the user should be authenticated or not.

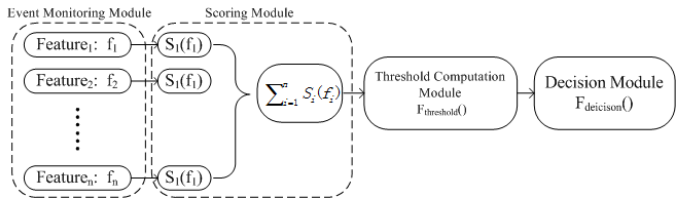

Fig. 1. System Block Diagram

The features used to model user profile can be represented as:

$$Features : (f_1, f_2, f_3, ..., f_n) \quad (1)$$

The functions to compute the score for each feature are represented as:

$$Functions : (S_1, S_2, S_3, ..., S_n) \quad (2)$$

Thus, the aggregate score $AS$ computed by this model can be represented as:

$$AS = \sum_{i=1}^{n} S_i(f_i) \quad (3)$$

Since the scheme is event driven, an aggregate score is calculated with each occurrence of a feature-related event. Hence, an aggregate score stream ($AS_1, ..., AS_k$) is obtained which is then utilized to compute the threshold. Thus:

$$(Threshold_1, ..., Threshold_k) = F_{threshold}(AS_1, AS_2, ..., AS_k) \quad (4)$$

where $F_{decision}$ denotes the algorithm used to compute a suitable decision threshold. After the threshold is computed, the system is able to make the authentication decision based on the current aggregate score and current threshold at any given time $t$ (assuming $t$ is between event$_i$ and event$_{i+1}$).

$$Decision = F_{decision}(AS_i, Threshold_i, t) \quad (5)$$

where $F_{decision}$ denotes the approach which the system employed to determine whether the user passes the authentication or not.

B. Scoring Algorithm

A user profile consists of a number of features which are employed to generate a final aggregate score. The algorithm developed to compute the score based on each feature is explained in the following.

*Incoming/Outgoing SMS*: Each time an incoming/outgoing SMS event occurs, the system records the time and a representation of the phone number. Each incoming/outgoing phone number is assigned a priority value calculated from:

$$Value = 120 + (48 \times OT) - t \quad (6)$$

where $OT$ denotes the total number of times the event has occurred, and $t$ is the time between the current computation and the last computation (in hours). With the above expression, a cache of the most relevant events is maintained, as any value less than zero is removed from the list. $OT$ is initialized to zero for any number removed from the list. Additionally, the formula allows for numbers that have been inactive for more than a week to be removed from the cache. Each time a new incoming/outgoing SMS event occurs, the system checks if the following conditions have been met whilst calculating the score $Score_{(is/os)}$:

*Condition$_1$*: The system searches if this number is in the top 5 positions in the priority value list.

*Condition$_2$:* The system checks if this number appears in the user's contact list.

The score is defined as follows:

$$Score_{(is/os)} = \sum Score_{condition} \quad (7)$$

*Incoming/Outgoing call*: The score is computed in the same way as incoming/outgoing SMS except that it also checks the duration of the phone call. A long duration call is an indication of normal usage and this is taken into consideration in the score computation.

*Browser history*: The system extracts the browser history and records the domain name of each visited URL. Similarly to the SMS and call features, a priority list is maintained based on the calculation (for each URL):

$$Value = 72 + (8 \times OT) - t \quad (8)$$

where $OT$ represents the total number of times the URL is visited and $t$ (hours) is the interval between the last visit and the current visit. A cache of the most relevant URLs is maintained by removing negative values from the list, and $OT$ for the removed URL is reset to zero. Note that browser history is extracted periodically (in this study, every 20 minutes) as there's no event notification that can be used in the system to easily detect browsing activity. Thus, it checks how many domains have been visited in previous 20 minutes. The score for browser history is computed by checking how many newly visited domains are in the top 6 positions in the priority value list. The score is defined as follow:

$$Score_{browser} = \sum Score_{domain} \quad (9)$$

**WIFI history**: The system records the service set identifier (SSID) of the connected access point and the duration of each connection. A priority value list is also computed similar to the previous features:

$$Value = 100 + (18 \times OT) - t \quad (10)$$

where $OT$ denotes the total number of times the user connected to a specific access point and $t$ (hour) denotes the interval between the last connection time and the current time. Connections that are less than two minutes are not included in the computation in order to avoid employing automatic short connections to access points on the move. Each time a new WIFI connection is established, the system determines the score for WIFI history by checking whether it is within the top 5 from the priority value list and then computes the score as follows:

$$Score_{ap} = \sum Score_{condition} \quad (11)$$

**Damping Factor:** A damping factor is introduced in the computation of the score for each feature in order to compensate for long periods of inactivity and allow for gradual decrease in the score. The damping factor $\mu$ is used in the computation of the score for each feature as follows:

$$Score = Score - \mu \times t \quad (12)$$

where $\mu$ can be adjusted to meet the requirement of different situations.

### C. Computing the decision threshold

After computing an aggregate score from the individual feature's scores, a decision threshold is determined below which the system infers abnormal behavior and above which a routine user behavior is assumed. An inference of abnormal behavior is then the trigger for activating extra layer(s) of authentication (e.g. explicit authentication.).

In our scheme, we propose the use of an adaptive threshold in order to track and respond to shifts in routine user behavior. Additionally, it is preferred to minimize the training requirements for the model. Hence, we develop and evaluate three threshold computation techniques: SD_Block, EWMA_Direct and EWMA_SD_Block. These are also compared to a static threshold based scheme that incorporates a training phase. The investigated decision threshold computation techniques are described as follows:

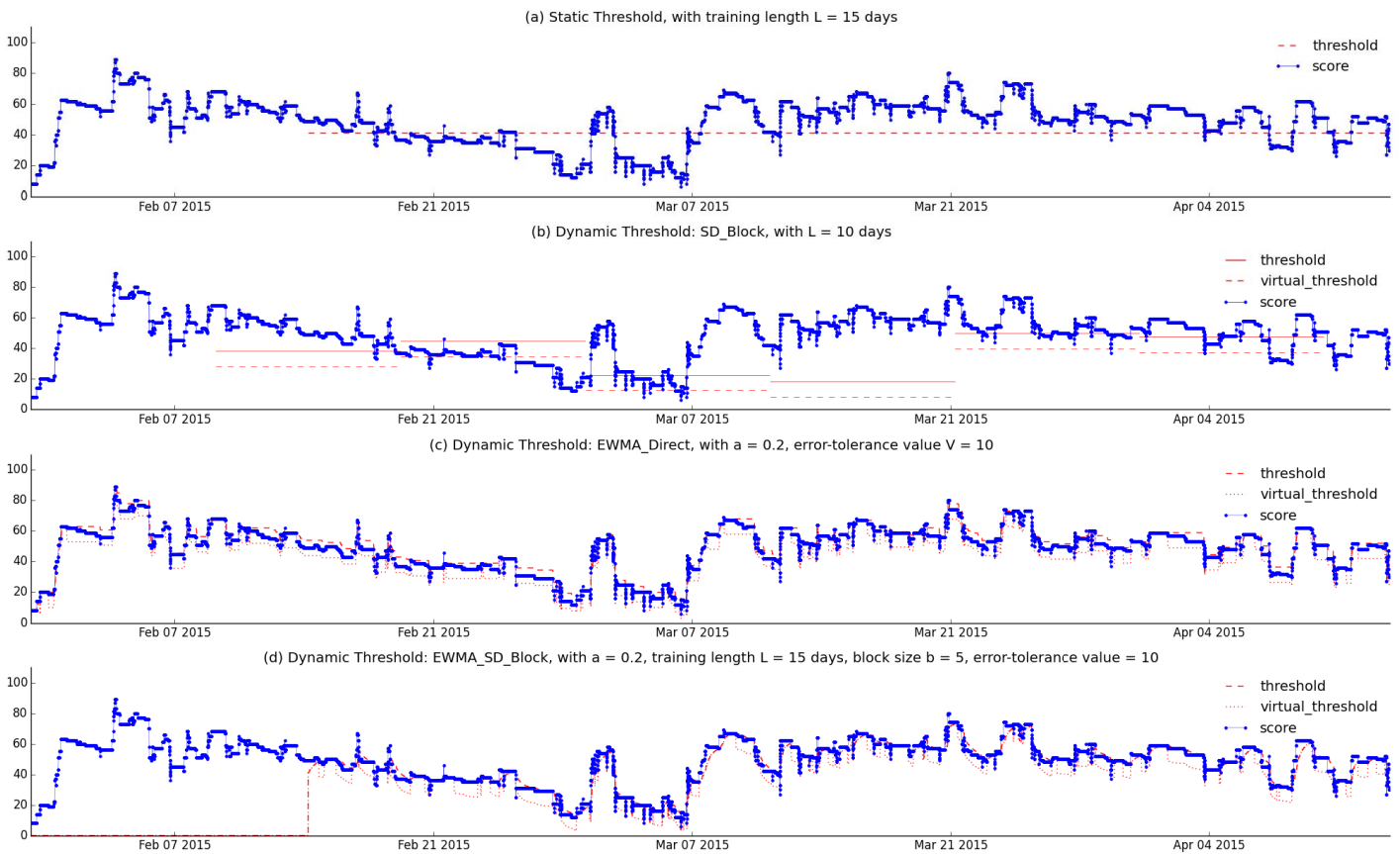

Fig. 2. Examples of different threshold techniques

*1) Static threshold computation*

In static threshold, when the training phase is finished, the system calculates mean value and standard deviation (of the aggregate scores) in order to compute the threshold which remains unchanged throughout the monitoring phase. Static threshold is illustrated in Fig. 2 (a) alongside aggregate scores computed from real data collected between February and April 2015 from a test user:

$$Threshold = mean - (standard\_deviation) \quad (13)$$

*2) Dynamic threshold computation*

The dynamic threshold computation techniques are based on standard deviation and EWMA (exponentially weighted moving average). EWMA is chosen for its potential to enable training free decision support.

*a) SD_Block*

This threshold computation technique is based on the standard deviation. The system divides the aggregate score stream into multiple equal-length blocks ($block_1, …, block_n$) of length $L$. The first block is used for training and the computed threshold is used within the second block. This pattern is repeated continuously (i.e. previous block provides training scores for computing threshold for current block as illustrated in Fig. 2 (b), alongside aggregate scores computed from real data collected between February and April 2015 from a test user) . An additional (virtual) threshold is also depicted which is derived from incorporating error tolerance in the original threshold computation. The details are discussed in section III D.

*b) EWMA_Direct*

EWMA method has been previously utilized by other scientific researches in a similar manner, such as admission control schemes [12] and buffer management [13, 14]. This method uses the weighted sum of previous thresholds and the current aggregate score output to compute the new threshold.

$$Threshold_1 = AS_1$$
$$Threshold_t = \alpha \cdot AS_{t-1} + (1-\alpha) \cdot Threshold_{t-1}, \quad t > 1 \quad (14)$$

Where $AS$ is the aggregate score described in the previous part, $\alpha$ is the weight which determines the importance of historical data in $Threshold_t$. The smaller $\alpha$ implies that the new threshold relies more on the previous thresholds and gradually draws close to the aggregate score. In the implementation, the $\alpha$ is set initially as 0.2 (as similarly applied in [6, 12]) since this is reasonable for the threshold adaptation speed in the system. In this method, the threshold is constantly changing whenever a feature-related event occurs, this can be clearly seen in the Fig. 2 (c). In addition, as shown in the figure, the proposed authentication scheme which uses EWMA_Direct threshold computation method needs no initial training period.

*c) EWMA_SD_Block*

This method is a hybrid of the SD_Block and EWMA_Direct. Firstly, it uses a block of output aggregate score ($block_1$ in SD_Block) as training dataset to compute the $threshold_1$. Secondly, it employs EWMA method to compute a new threshold adaptively as outlined in EWMA_Direct.

$$Threshold_1 = mean - (standard\_deviation) \quad (15)$$
$$Threshold_t = \alpha \cdot ASBA_{t-1} + (1-\alpha) \cdot Threshold_{t-1}, \quad t > 1$$

Here, $ASBA$ stands for Aggregate Score Block Average. Instead of using single total score value as input to $ASBA$, the system uses the average of the current block (block size $b$ can be adjusted in the implementation) as input to $ASBA$. The new computed threshold is used as decision threshold for the next block. For example, assuming initial $AS$ (aggregate score) index of current block is $k$, block size is set by $b$. The value of $ASBA$ is expressed as follow:

$$ASBA = \frac{\left(\sum_{j=k}^{j=k+b} AS_i\right)}{b} \quad (16)$$

And the new computed threshold is used as detection threshold for next upcoming block from $AS_{k+b+1}$ to $AS_{k+2b+1}$. EWMA_SD_Block threshold computation is shown in Fig. 2 (d) alongside aggregate scores computed from real data collected between February and April 2015 from a test user

*D. Error Tolerance*

An error tolerant design is incorporated to the decision threshold detection in order to allow for tuning to improve user recognition rates. This allows for a margin of error to be made in the decision while still being able to detect unusual behavior that would indicate adversarial use in a theft scenario. It considers a value V such that when the score is below the threshold but by less than V, a normal user behavior is inferred. Thus, term 'virtual threshold' is introduced into the system:

$$Virtual\_threshold = threshold - V$$
$$AS > threshold, \quad normal$$
$$Virtual\_threshold \le AS < threshold, \quad normal$$
$$AS < Virtual\_threshold, \quad abnormal \quad (17)$$

$V$ can be adjusted based on different user requirements. For example, if users want the system to be able to accommodate more false alarms, they can keep $V$ at a high value. If this is not acceptable from a security point of view, they can keep $V$ at a low value to accommodate more explicit authentication. The table below gives the user recognition rate before and after fault tolerance is implemented, the rest of parameters use configuration illustrated in Fig. 1. As it is shown in Table II, a dramatic increase of *user recognition rate* is observed after fault tolerance is deployed into the system, with EWMA_Direct increasing from 64.18% to 98.36% and EWMA_SD_Block increasing from 54.98% to 95.85%.

TABLE II. USER RECOGNITION RATE BEFORE AND AFTER ERROR TOLERANCE IS DEPLOYED

| | SD_Block | EWMA_Direct | EWMA_SD_Block |
|---|---|---|---|
| Error Tolerance (on V = 10) | 98.28% | 98.36% | 95.85% |
| Error Tolerance (off V = 0 ) | 81.12% | 64.18% | 54.98% |

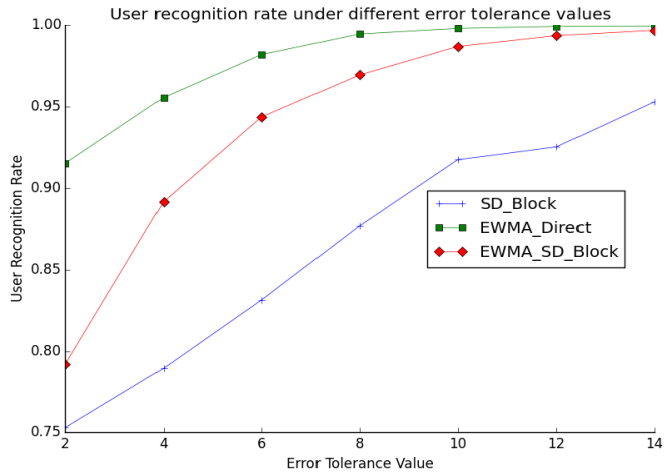

Fig. 3. The impact of error tolerance value on user recognition rate

## IV. EVALUATION

This section presents the impact different parameters have on the user recognition rate and presents the performance of the proposed scheme in adversary test scenarios (i.e. simulated device theft). An Android application is developed to extract the data based on the features described previously. In order to make the evaluation easier, the scheme and threshold computation techniques are implemented using Python scripts to emulate the logic of the authentication scheme.

### A. User recognition rate

Fig. 3 to 5 depict the results of the experiments undertaken to evaluate the accuracy of the developed authentication scheme. Data is obtained from a Samsung Galaxy A5 device with Android version 4.4.4. The developed application requires a minimal Android SDK version of 8.

Fig. 3 shows an ascending user recognition rate for SD_Block, EWMA_Direct and EWMA_SD_Block threshold computation methods when the error tolerance value is gradually increased. As mentioned earlier, a higher error tolerance value indicates that the authentication scheme can reduce false alarm rate.

Fig. 4 depicts how training length can affect the user recognition rate in static threshold, SD_Block and EWMA_SD_Block. From Fig. 4(a), a high performance from static threshold is observed from training for 1 to 3 days, and then drops dramatically afterward. This is because the score at the beginning of training is relatively low compared with other periods (score for some features is 0 if the user has done nothing related to that feature). Fig. 4(a) also indicates that dynamic threshold performs much better than the static threshold. This is because user behavior is changing constantly and hence requires the threshold to react adaptively in the same way, which the dynamic threshold scheme is capable of.

Fig. 4(b) is the magnification of Y axis between SD_Block and EWMA_SD_Block in Fig. 3(a), it shows that EWMA_SD_Block performs much better than the SD_Block. An interesting observation is that the training length barely has impact on user recognition rate of EWMA_SD_Block which fluctuates around 98% user recognition rate.

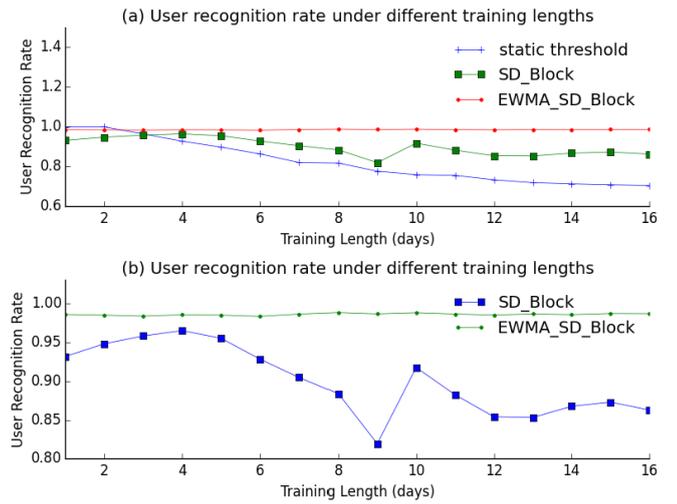

Fig. 4. The impact of training length on user recognition rate

EWMA_Direct threshold method does not require training phase (can be seen in Fig. 2(c)), which implies that it can be deployed to detect abnormal behavior shortly after it is installed on the users' mobile devices.

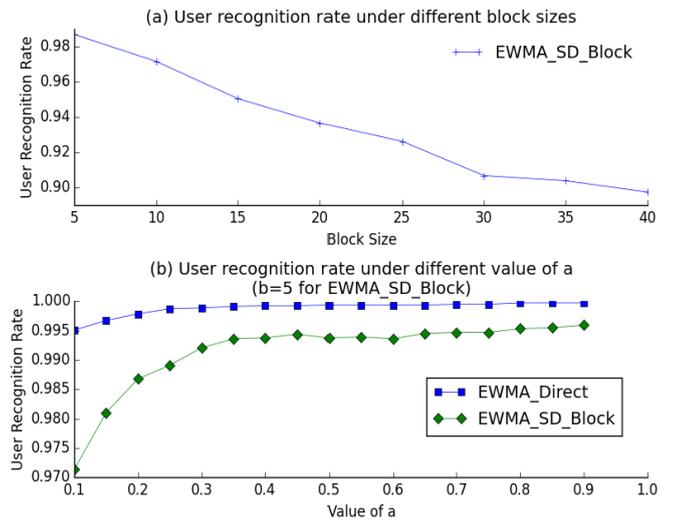

Fig. 5. The impact of block size and α value on user recognition rate

Fig. 5(a) illustrates that the major trend of user recognition rate roughly decreases with the increasing block size. This is due to the EWMA algorithm which gives the threshold more space to catch up with the change of the score when the block size is smaller. The result indicates that small-sized block can achieve a better performance in user recognition.

Fig. 5(b) shows an ascending trend of user recognition rate with α value increasing. In EWMA, the coefficient α represents the degree of weighting, a higher α discounts older observation (threshold) faster. Hence, high value of α implies the new threshold computation is based more upon the current score instead of the previous threshold. In reality low α value is expected so that the new computed threshold changes steadily and gradually draws close to the current aggregate score.

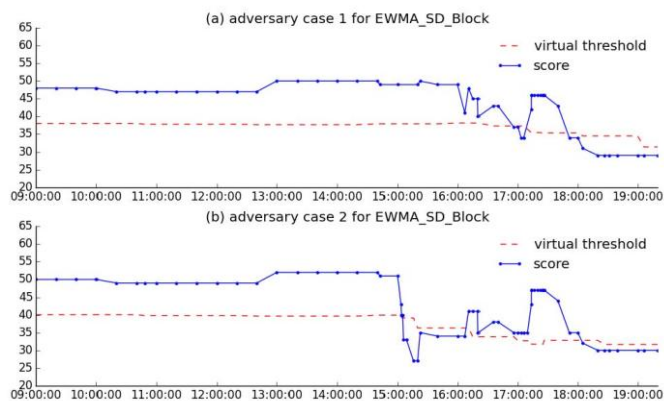

Fig. 6. Example of adversary case 1 and 2

*B. Adversary case study*

Two adversary cases were investigated to emulate device theft and the results are presented in Fig. 6. The attack starts at 15:00:00 (the device is taken by the attacker). The performance are evaluated by observing two metrics: *NOC* (number of computations before the first detection as illegitimate use) and *elapsed time* in minutes (before first detection).

*Case 1:* The attacker takes the device and interacts with it continuously but not intensively. E.g. connecting to WIFI previously unknown to the device, browsing unfamiliar websites, calling/texting numbers unfamiliar to the device.

*Case 2:* This case represents a more intensive and immediate interaction with the device after it is taken from the legitimate user.

TABLE III    PERFORMANCE IN ADVERSARY CASE 1

|  | NOC | Elapsed time(minutes) |
|---|---|---|
| Static threshold | N/A | N/A |
| SD_Block | 29 | 200 |
| EWMA_Direct | 26 | 171 |
| EWMA_SD_Block | 12 | 115 |

TABLE IVI    PERFORMANCE IN ADVERSARY CASE 2

|  | NOC | Elapsed time(minutes) |
|---|---|---|
| Static threshold | N/A | N/A |
| SD_Block | 9 | 16 |
| EWMA_Direct | 2 | 3 |
| EWMA_SD_Block | 7 | 5 |

Table III and IV indicate that case 2 detects the non-legitimate use much faster than case 1. This is because the proposed authentication scheme is event driven and case 2 has more intensive activities on the device. Table III and IV also shows that static threshold is unable to detect the illegitimate use and EWMA-based threshold method performs better than SD_Block. EWMA_Direct shows a relatively better performance than EWMA_SD_Block in adversary case 2 while the latter performs better in case 1.

The impact different parameters have on NOC and elapsed time is also evaluated during the experiment. Due to space limit, the results shown in Fig. 7 to 9 are for only adversary case 1.

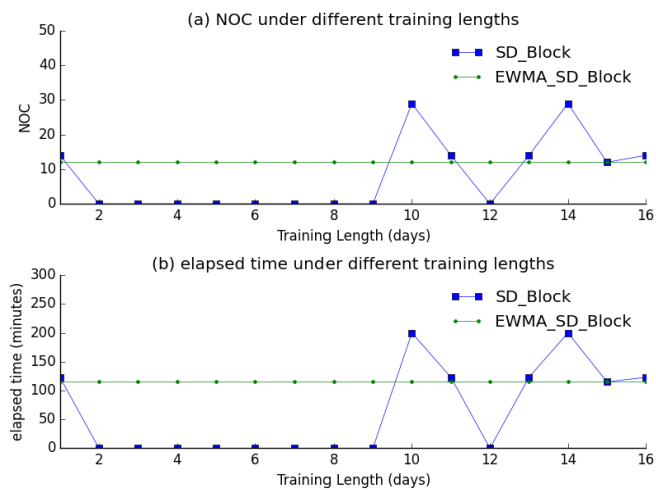

Fig. 7. The impact of training length on NOC and elapsed time

An interesting observation in Fig. 7 is that NOC and elapsed time of EWMA_SD_Block are independent of the training length. Considering also the result from Fig. 4, we can conclude that EWMA_SD_Block requires very minimal or no training as well. Meanwhile, considering the user recognition rate for SD_Block in Fig. 4(b), the optimal value of training length for SD_Block could be determined in our experiment, which is between 2 days and 6 days.

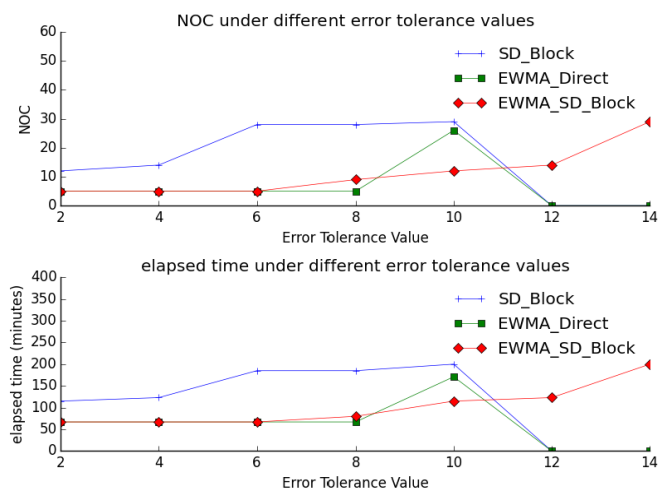

Fig. 8. The impact of error tolerance value on NOC and elapsed time

Fig. 8 illustrates the change of NOC and elapsed time under different error tolerance values. Considering the results from both Fig. 3 and 8, it can be concluded that the optimal value for error tolerance is between 8 and 10 for SD_Block, and between 6 and 8 for EWMA_Direct, and between 6 and 10 for EWMA_SD_Block. Fig. 9 indicates that smaller value of block size can improve the performance of EWMA_SD_Block based authentication scheme on illegitimate use detection.

Combining with results obtained from Fig. 5(a), it can be concluded that the optimal value for block size *b* is between 6 and 10. These optimal values are the best operating points for the scheme that can simultaneously enable the highest user recognition rates and quickest adversary scenario detection (i.e. lowest values for NOC and elapsed time).

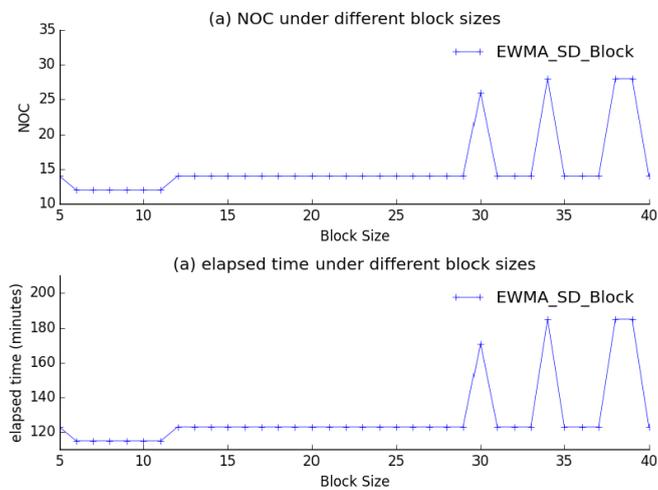

Fig. 9. The impact of block size on NOC and elapsed time

## V. CONCLUSION

This paper proposed and evaluated an event-driven implicit authentication scheme for mobile access control. The scheme is designed to operate transparently and adaptively by computing aggregate scores (indicative of the level of trust of the current user) which are compared to dynamic decision thresholds to enable the recognition of legitimate user activities while detecting adversary use cases. The proposed authentication scheme also requires no user interaction, and can be used to trigger explicit authentication to prevent unauthorized access in adversarial scenarios. Experiments were conducted to investigate the performance of the schemes using real data collected from a test user (during normal routine behavior) between February and April 2015. Afterwards, two adversary use cases were emulated to measure the effectiveness of illegitimate use detection.

The results show that it is feasible and practical to build an effective implicit authentication system from readily available and easily obtainable everyday phone data. Future work will explore a more extensive range of data sensors and other possible behavior profiling algorithms. Furthermore, a full prototype of the authentication system will be implemented and evaluated on Android.